\documentstyle[editedvolume]{crckapb} 
\input tex.def
\input psfig.tex

\newcommand{\farcs}{\hbox{$.\!\!^{\prime\prime}$}}

\begin{opening}
\title{SUPERMASSIVE BLACK HOLES IN GALACTIC NUCLEI}
\subtitle{Observational Evidence and Some Astrophysical 
Consequences\footnote{Invited review paper to appear in {\it Observational 
Evidence for Black Holes in the Universe}, ed. S.~K. Chakrabarti (Dordrecht: 
Kluwer).}} 
\author{LUIS C. HO}
\institute{Harvard-Smithsonian Center for Astrophysics\\
60 Garden St., Cambridge, MA 02138, USA}
\end{opening}

\begin{document}

\begin{abstract}

I review the status of observational determinations of central masses in 
nearby galactic nuclei.  Results from a variety of techniques are summarized, 
including ground-based and space-based optical spectroscopy, radio VLBI 
measurements of luminous water vapor masers, and variability 
monitoring studies of active galactic nuclei.  I will also discuss recent
X-ray observations that indicate relativistic motions arising from the
accretion disks of active nuclei.  The existing evidence suggests that 
supermassive black holes are an integral component of galactic structure, 
at least in elliptical and bulge-dominated galaxies.  The black hole mass 
appears to be correlated with the mass of the spheroidal component of the 
host galaxy.  This finding may have important implications for many 
astrophysical issues.

\end{abstract}

\section{Motivation}

The discovery of quasars in the early 1960's 
quickly spurred the idea that these amazingly powerful sources derive their 
energy from accretion of matter onto a compact, extremely massive object, 
most likely a supermassive black hole (SMBH; Zel'dovich \& Novikov 1964; 
Salpeter 1964; Lynden-Bell 
1969) with $M\,\approx\,10^6-10^9$ \solmass.  Since then this model has 
provided a highly useful framework for the study of quasars, or 
more generally, of the active galactic nucleus (AGN) phenomenon (Rees 1984; 
Blandford \& Rees 1992).  Yet, despite its success, there is little empirical 
basis for believing that this model is correct.  As pointed out by Kormendy 
\& Richstone (1995, hereafter KR), our confidence that SMBHs must power AGNs 
largely rests on the implausibility of alternative explanations.  To be sure, 
a number of characteristics of AGNs indicate that the central engine must be 
tiny and that relativistic motions are present.  These include 
rapid X-ray variability, VLBI radio cores, and superluminal motion.  However, 
solid evidence for the existence of SMBHs in the centers of galaxies has, 
until quite recently, been lacking.

As demonstrated by Soltan (1982), simple considerations of the quasar 
number counts and standard assumptions about the efficiency of energy 
generation by accretion allows one to estimate the mean mass density of SMBHs 
in the universe.  The updated analysis of Chokshi \& Turner (1992) finds 
$\rho_{\bullet}\,\approx\,2 \times 10^5 \epsilon_{0.1}^{-1}$ 
\solmass\ Mpc$^{-3}$ for a radiative efficiency of 
$\epsilon\,=\,0.1\epsilon_{0.1}$.  Comparison of $\rho_{\bullet}$ 
with the $B$-band galaxy luminosity density of 
1.4\e{8}$h$ \solum\ Mpc$^{-3}$ (Lin \etal 1996), where the Hubble constant 
$H_{\rm 0}$ = 100$h$ \kms\ Mpc$^{-1}$, implies an average SMBH mass per unit 
stellar luminosity of $\sim$1.4\e{-3}$\epsilon_{0.1}^{-1} h^{-1}$ 
\solmass/\solum.  A typical bright galaxy with $L_B^*\,\approx\,10^{10} 
h^{-2}$ \solum\ potentially harbors a SMBH with a mass
\gax $10^7\epsilon_{0.1}^{-1}h^{-3}$ \solmass.  These very general 
arguments lead one to conclude that ``dead'' quasars ought to be lurking 
in the centers of many nearby luminous galaxies.

The hunt for SMBHs has been frustrated by two principal limitations.  The 
more obvious of these can be easily appreciated by nothing that the ``sphere 
of influence'' of the hole extends to $r_{\rm h}\,\simeq\,G M_{\bullet}/
\sigma^2$ (Peebles 1972; Bahcall \& Wolf 1976), where $G$ is the gravitational 
constant and $\sigma$ is the velocity dispersion of the stars in the bulge, 
or, for a distance of $D$, $\sim$1\asec
($M_{\bullet}/2\times10^8$ \solmass)($\sigma$/200 \kms)$^{-2}$($D$/5 Mpc).  
Typical ground-based observations are therefore severely hampered by 
atmospheric seeing, and only the heftiest dark masses in the closest galaxies 
can be detected.  The situation in the last few years has improved dramatically 
with the advent of the {\it Hubble Space Telescope (HST)} and radio VLBI 
techniques.  The more subtle complication involves the actual modeling
of the stellar kinematics data, and in this area much progress has also 
been made recently as well.

Here I will highlight some of the observational efforts during the past 
two decades in searching for SMBHs, concentrating on the recent advances.  
Since this contribution is the only one that discusses nuclear BHs
aside from that in the Milky Way (Ozernoy, these proceedings) and in 
NGC 4258 (Miyoshi, these proceedings), I will attempt 
to be as comprehensive as possible, although no claim to completeness is made, 
as this is a vast subject and progress is being made at a dizzying pace.  To 
fill in the gaps, I refer the reader to several other recent review papers, 
each of which has a slightly different emphasis (KR; Rees 1998; 
Richstone 1998; Ford \etal 1998; van der Marel 1998).

\section{Early Clues from Photometry}

The prospect of finding massive BHs in globular clusters motivated 
much early effort to investigate the distribution of stars resulting from the 
adiabatic growth of a BH in a preexisting stellar system.  The central 
density deviates strongly from that of an isothermal core and instead follows 
a cuspy profile $\rho(r)\,\propto\,r^{-3/2}$ (Young 1980) or 
steeper if two-body relaxation (Peebles 1972; Bahcall \& Wolf 1976)
or different initial density profiles (Quinlan, Hernquist, \& 
Sigurdsson 1995) are taken into account.  The discovery that the centers of 
some giant elliptical galaxies obey this prediction 
generated much enthusiasm for the existence of SMBHs.  In the 
well-known case of M87 (Young \etal 1978), Lauer \etal (1992) have since shown 
that the central cusp persists to the limit of the resolution of the {\it HST} 
(0\farcs1).  

However, as emphasized by Kormendy (1993; see also KR), photometric signatures 
alone do not uniquely predict the presence of a SMBH.  The cores of most 
galaxies are now known to be nonisothermal. And moreover, contrary to 
na\"\i ve expectations, galaxy cores with high central surface brightnesses and 
small core radii, far from being the ones most likely to host SMBHs, are 
in fact {\it least} expected to do so.  This apparently contradictory 
statement can be most easily understood by considering the so-called 
fundamental-plane relations for the spheroidal component of galaxies 
(Faber \etal 1987; Bender, Burstein \& Faber 1992).  More luminous, more 
massive galaxies tend to have more massive central BHs (\S\ 7), but 
they also have larger, more diffuse cores.
Indeed, high-resolution photometric studies of early-type 
galaxies (Nieto \etal 1991; Crane \etal 1993; Jaffe \etal 1994; Lauer \etal 
1995) find that the central surface brightness profiles either continue 
to rise toward the center as $I(r)\,\propto\,r^{-\gamma}$, with 
$\gamma\,\approx$ 0.5--1.0 (the ``power-law'' galaxies) or they flatten at 
some characteristic radius to a shallower slope of $\gamma\,\approx$ 0.0--0.3
(the ``core'' galaxies).  The power-law galaxies are invariably lower 
luminosity, lower mass systems compared to those with distinct cores.

In summary, photometric signatures alone cannot be used as reliable indicators
for the presence of SMBHs.  Instead, we must turn to the more arduous task 
of obtaining kinematic measurements.

\section{Methods Based on Stellar Kinematics}

Contrary to the ambiguity of light profiles, the Keplerian rise in the 
velocity dispersion toward the center, $\sigma(r)\,\propto\,r^{-1/2}$, is a 
robust prediction for a wide variety of dynamical models containing a 
central massive dark object (MDO; 
Quinlan \etal 1995).  Sargent \etal (1978) noticed that the innermost 
velocities of M87 were consistent with such a prediction, and, assuming an 
isotropic velocity distribution, they inferred that the center of this galaxy 
contained a dark mass of $\sim$5\e{9} \solmass, presumably in the form of a
SMBH.  The central rise in $\sigma(r)$, unfortunately, can be insidiously 
mimiced by an anisotropic velocity distribution, and therefore an MDO is 
{\it not} required by the data for this object (Duncan \& Wheeler 1980; Binney 
\& Mamon 1982; Richstone \& Tremaine 1985; Dressler \& Richstone 1990; van der 
Marel 1994a). This degeneracy presents a serious difficulty for many mass 
determinations based on stellar kinematic data.  An extensive and lucid 
discussion of this vast subject was presented by KR, and many of the 
details will not be repeated here.  Nonetheless, an abbreviated synopsis 
is needed to motivate the topic.

Following the notation of KR, the radial variation in mass 
can be expressed by the first velocity moment of the collisionless 
Boltzman equation,

\begin{displaymath}
M(r) = {{V^2 r}\over{G}} 
+ {{\sigma_r^2 r}\over{G}}\bigg[
- {{{\rm dln}\, \nu}\over{{\rm dln}\, r}} 
- {{{\rm dln}\, \sigma_r^2}\over{{\rm dln}\, r}} 
- (1 - {{\sigma_{\theta}^2}\over{\sigma_r^2}}) 
- (1 - {{\sigma_{\phi}^2}\over{\sigma_r^2}})\bigg]
\end{displaymath}

\noindent where $V$ is the rotational velocity, $\sigma_r$ is the radial and
$\sigma_{\theta}$ and $\sigma_{\phi}$ the azimuthal components of the 
velocity dispersion, and $\nu$ is the density of the tracer population.  
In practice, several simplifying assumptions are adopted: (1) the mass 
distribution is spherically symmetric; (2) the mean rotation is circular; 
and (3) $\nu$ is proportional to the luminosity density, or, 
equivalently, that $M/L$ does not vary with radius.  

A brief scrutiny of the above equation indicates that the effects of velocity 
anisotropy can have a large and complicated effect on the derivation of $M(r)$ 
because the terms inside the bracket significantly affect the 
${\sigma_r^2 r}/G$ term.  If 
$\sigma_r\, >\, \sigma_{\theta}$ 
and $\sigma_r\, >\,\sigma_{\phi}$, each of the last two terms will be negative 
and can be as large as --1.  The central brightness distributions of the 
spheroidal component of most galaxies typically have 
$-({\rm dln}\, \nu/{\rm dln}\, r)\,\approx\,+1.1$ for luminous, nonrotating 
systems and \gax +2 for low to intermediate-luminosity systems (e.g., 
Faber \etal 1997).  Since $-({\rm dln}\, \sigma_r^2/{\rm dln}\, r)\,\leq\,+1$, 
it is apparent that, under suitable conditions, all four terms can largely 
cancel one another.  As emphasized by KR, all else being equal, smaller, 
lower luminosity galaxies such as M32 potentially yield more secure mass 
determinations than massive galaxies like M87 because less luminous systems 
tend to have (1) steeper central light profiles, (2) a greater degree of 
rotational support, and (3) less anisotropy.

The principles behind the stellar kinematics analysis are conceptually 
straightforward but in practice technically challenging.  Given the 
set of observed quantities $I(r)$, $V(r)$, and $\sigma(r)$, the goal is to 
derive a range intrinsic values for these quantities after accounting for 
projection and the blurring effects of seeing.  Much of the machinery for 
these tasks has been developed and extensively discussed by Kormendy 
(1988a, b) and Dressler \& Richstone (1988).  The sensitivity of the 
results to the effects of anisotropy are examined through maximum-entropy 
dynamical models (Richstone \& Tremaine 1984, 1988) to 
see whether conclusions regarding the presence of MDOs can be obviated
by a suitable exploration of parameter space.  Perhaps the most serious 
limitation of these maximum-entropy models is that they do not properly take 
flattening into account.

The last several years have seen a resurged interest in improving the 
techniques of analyzing stellar kinematics data.  In the context of SMBH 
searches, Gerhard (1993), van der Marel \etal (1994a, b), Dehnen (1995), 
among others, have stressed the importance of utilizing the full 
information contained in the velocity profile or line-of-sight velocity 
distribution (LOSVD) of the absorption lines, which 
are normally treated only as Gaussians.  A system with significant rotation, 
for instance, can leave a measurable skewness on the LOSVD, while 
various degrees of anisotropy would imprint symmetric deviations from a
Gaussian line shape.  Neglecting these subtleties can lead to systematic 
errors in the measurement of $V(r)$, but in the cases best studied so far 
these effects do not seem to have been severe (KR).  Furthermore, the 
line profile should develop weak, high-velocity wings if a SMBH is present 
(van der Marel 1994b), although the currently available data do not yet have 
the requisite quality to exploit this tool.  

Yet another advance has focused on the development of dynamical models with 
two-integral phase-space distribution functions, $f(E,L_z)$, $E$ being the 
total energy and $L_z$ the angular momentum in the symmetry axis (van der Marel 
\etal 1994b; Qian \etal 1995; Dehnen 1995).  Such models are properly 
flattened, and they generate predictions for the LOSVDs; on the 
other hand, it is not clear whether imposing a special dynamical structure 
is too restrictive.  This limitation will be eliminated by fully general, 
axisymmetric three-integral models (van der Marel \etal 1998;
Cretton \etal 1998; Gebhardt \etal 1998). 

There are currently 10 galaxies with published MDO measurements determined
from stellar kinematical data (Table 1).  Of these, only three (M81: 
Bower \etal 1996; NGC 3379: Gebhardt \etal 1998; NGC 4342: van den Bosch 1998) 
come solely from {\it HST} data; the remaining ones, although 
many by now confirmed with {\it HST}, were initially discovered 
from high-quality ground-based observations (see KR for a detailed account 
of each object).  Kormendy and collaborators, in particular, making use of 
the excellent seeing conditions and instrumentation on the CFHT, continue to 
make progress in this area.  Two new MDOs have been reported recently based 
on CFHT data: the low-luminosity elliptical galaxy NGC 4486B has 
$M_{\rm {\small{MDO}}}$ = 6\e{8} \solmass\ (Kormendy \etal 1997b), and NGC 
3377, another close cousin, has $M_{\rm {\small{MDO}}}$ = 2.3\e{8} 
\solmass\ (Kormendy \etal 1998).  This demonstrates the important fact that 
even in the {\it HST} era ground-based observations continue to play an 
important role in SMBH searches.

\begin{figure}   
\hskip 0.75truein
\psfig{file=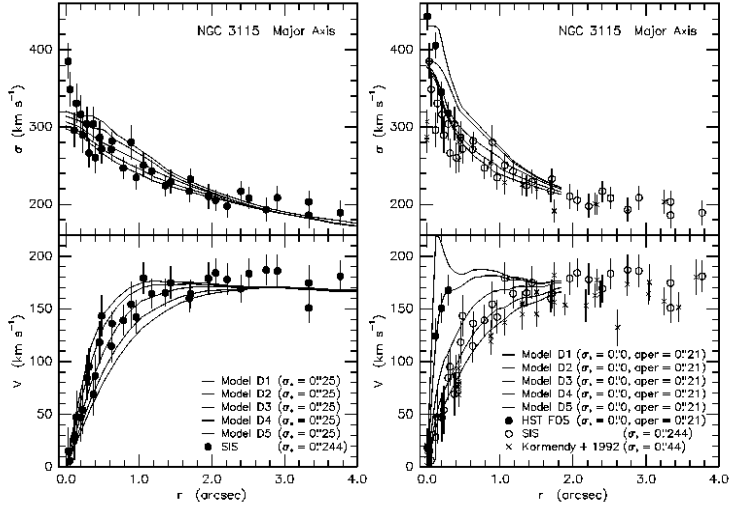,width=3.0truein,angle=0}
\caption{Stellar kinematic data for NGC 3115 compared with various
dynamical models (see Kormendy \etal 1996).  The {\it left} panel shows
the best ground-based data, and the {\it right} panel the same data with
new {\it HST} data superposed for comparison.  Both $V$ and $\sigma$ rise
much more steeply with radius in the new {\it HST} data.}
\end{figure}

The new observations with {\it HST}, thus far all acquired using the 
Faint Object Spectrograph (FOS), provide an important contribution by 
increasing the angular resolution by about a factor of 5 compared to the 
best ground-based data available.  In all cases studied (NGC 3115: 
Kormendy \etal 1996; NGC 4594: Kormendy \etal 1997a; M32: van der Marel \etal 
1997; M31: Ford \etal 1998), the velocity dispersions continue to rise toward 
smaller $r$ and the maximum rotational velocity has generally increased 
(Fig. 1).  In the case of NGC 3115, the FOS spectra are of sufficient 
quality to reveal wings in the LOSVD that extend up to $\sim$1200 \kms\ 
(Kormendy \etal 1996).  The {\it HST} data thus considerably bolster the case 
for a MDO in these objects.  The improvement in angular resolution additionally 
strengthens our confidence that the MDOs might indeed be SMBHs.  A reduction 
of the size scale by a factor of 5 increases the central density by more than 
two orders of magnitude.  Although in general this is still not enough to 
rule out alternative explanations for the dark mass (\S\ 6), it is 
clearly a step in the right direction.  

\begin{figure}
\hskip 1.0truein
\psfig{file=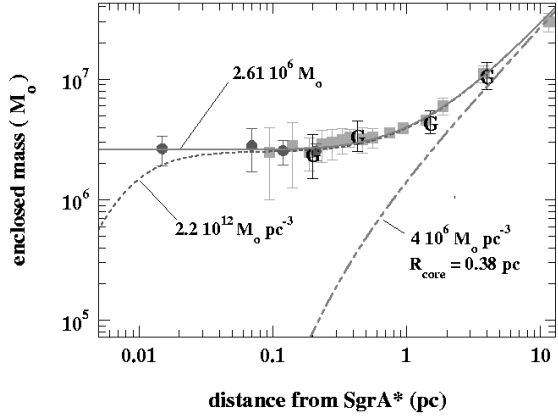,height=2.5truein,angle=0}
\caption{Enclosed mass versus radius for the Galactic Center derived from
stellar radial velocities and proper motions (from Genzel \etal 1997).
The points labeled with ``G'' come from gas kinematics.  The {\it thick
dashed} curve is a model for the stellar cluster with a total mass density of
4\e{6} \solmass\ pc$^{-3}$; the {\it solid} curve denotes the sum of this 
cluster and a point mass of 2.61\e{6} \solmass; and {\it thin dashed} curve 
is the sum of the stellar cluster and an additional dark cluster with a 
central density of 2.2\e{12} \solmass\ pc$^{-3}$.}
\end{figure}

I conclude this section with a few remarks on the dark mass in 
the Galactic Center (see Ozernoy in these proceedings for more details), 
which, in my view, is now the most compelling case of a SMBH in any 
galactic nucleus.  From analysis of an extensive set of near-IR radial 
velocities of individual stars, coupled with 
additional measurements from the literature, Genzel \etal (1996; see also 
Krabbe \etal 1995) found a highly statistically significant rise in the radial 
velocity dispersion between 5 and 0.1 pc from the dynamical center.  Assuming
an isotropic velocity distribution, the observations require a dark mass 
of $\sim$3\e{6} \solmass\ within $r$ = 0.1 pc and $M/L_K\,\geq\,100$;
the dark mass must have a density in excess of 10$^9$ \solmass\ 
pc$^{-3}$, which argues strongly for it being a SMBH.  These conclusions, 
and a suspicion nearly three decades old (Lynden-Bell \& Rees 1971),
have finally been vindicated by recent measurements of stellar proper motions 
within the central 1 pc region using high-resolution $K$-band astrometric 
maps (Eckart \& Genzel 1996, 1997; Genzel \etal 1997; Ghez \etal 1998).  
The main results are the following:  (1) the stellar radial velocities agree 
with the proper motions, which implies that on average the velocities are 
close to isotropic; (2) the combined velocities imply a dark mass (Fig. 2) 
within 0.006 pc of 2.61\e{6} \solmass\ (Genzel \etal 1997 quote a statistical 
error of $\pm$0.15 and a combined statistical and systematic error of 
$\pm$0.35); (3) the density, therefore, has an astonishingly high value of 
$>$2\e{12} \solmass\ pc$^{-3}$, which leaves almost no room to escape the 
conclusion that the dark mass must be in the form of a SMBH (\S\ 6).  The 
presence of a large mass is also supported by the detection of several stars, 
within 0.01 pc from the central radio source Sgr~A$^*$, moving at speeds in 
excess of 1000 \kms. From the velocities of the fast-moving stars and the near 
stationarity of Sgr~A$^*$, Genzel \etal further use equipartition 
arguments to constrain the mass of the radio core itself ($\geq 10^5$ 
\solmass), which, when combined with the exceedingly small upper limit 
for its size ($r\,<$ 4\e{-6} pc), would imply a density of 
$>$3\e{20} \solmass\ pc$^{-3}$.

\section{Methods Based on Gas Kinematics}

Unlike the situation for stars, gas kinematics are much easier to interpret 
if the gas participates in Keplerian rotation in a disklike configuration.  But 
there are two caveats to remember.  First, gas can be easily perturbed by 
nongravitational forces (shocks, radiation pressure, winds, magnetic fields, 
etc.).  Indeed, in the case of the Galactic Center, it was precisely this 
reason that its central mass, which had been estimated for some time using 
gas velocities (Lacy \etal 1980), could not be accepted with full confidence 
prior to the measurement of the stellar kinematics.  Second, there is 
no {\it a priori} reason that the gas should be in dynamical equilibrium, 
and therefore one must verify empirically that the velocity field indeed 
is Keplerian.  The optically-emitting ionized gas in the central regions of 
some spirals show significant noncircular motions (e.g., Fillmore, Boroson, 
\& Dressler 1986).  NGC 4594 is a striking example.  Kormendy \etal (1997a) 
showed that the emission-line rotation curve near the center falls 
substantially below the circular velocities of the stars, and 
hence the gas kinematics cannot be used to determine the central mass.

\subsection{Optical Emission Lines}

The sharpened resolution of the refurbished {\it HST} has revealed many 
examples of nuclear disks of dust and ionized gas (Fig. 3).  The nuclear disks 
typically have diameters $\sim$100--300 pc, with the minor axis often 
aligned along the direction of the radio jet, if present.  Some examples 
include the elliptical galaxies NGC 4261 (Jaffe \etal 1993), M87 (Ford \etal 
1994), NGC 5322 (Carollo \etal 1997), and NGC 315 (Ho \etal 1998), and the 
early-type spiral M81 (Devereux, Ford, \& Jacoby 1997).  I will highlight here 
only three cases; Table 1 gives a complete list of objects and references.

\begin{figure}
\hskip 0.0truein
\psfig{file=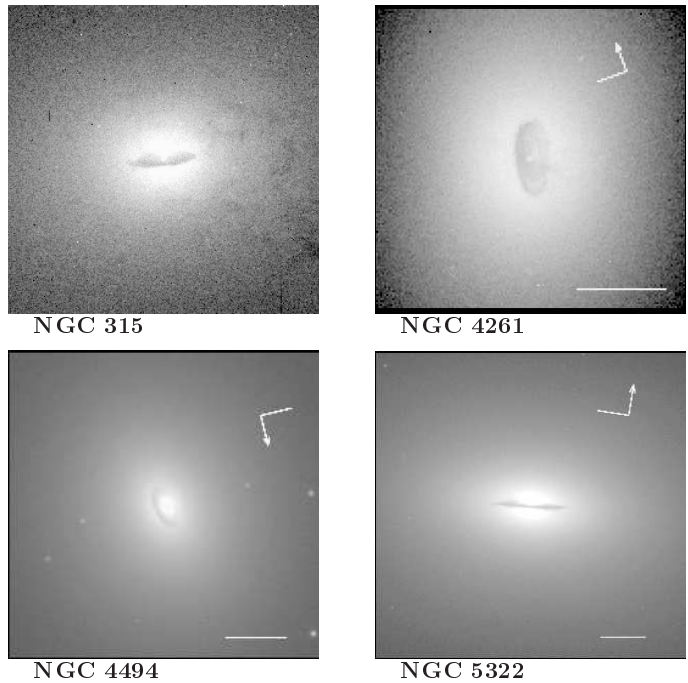,width=4.3truein,angle=0}
\caption{Nuclear disks from {\it HST} optical images. 
Each image is 35\asec\ on a side.} 
\end{figure}

The first object for which the nuclear gas disk was used to determine the 
central mass was M87.  Harms \etal (1994) used the FOS
to obtain spectra of several positions of the disk and 
measured a velocity difference of $\pm$1000 \kms\ at a radius of 0\farcs25 (18 
pc) on either side of the nucleus.   Adopting an inclination angle of 42\deg\ 
determined photometrically by Ford \etal (1994), the velocities were 
consistent with Keplerian motions about a central mass of 
(2.4$\pm$0.7)\e{9} \solmass.  Since the implied 
$M/L_V\,\approx$ 500, Harms et al. concluded that the central mass is 
dark, most likely in the form of a SMBH.  The case for a SMBH in M87 has 
been considerably strengthened through the recent reobservation with 
{\it HST} by Macchetto \etal (1997), who used the long-slit mode of the 
Faint Object Camera to obtain higher quality spectra extending to $r$ = 
0\farcs05 (3.5 pc).  The velocities in the inner few tenths of an arcsecond 
 are well fitted by a model of a thin disk in Keplerian rotation (Fig. 4),
although the inclination angle is not well constrained (47\deg\--65\deg).  
The rotation curve at larger radii falls below the Keplerian curve, possibly 
indicating a warp in the disk (Macchetto \etal 1997) or substantial 
perturbations due to spiral shocks (Chakrabarti 1995).  For $i$ = 52\deg, 
$M_{\rm {\small{MDO}}}$ = (3.2$\pm$0.9)\e{9} \solmass, and $M/L_V$ \gax 110.  
If, instead, a Plummer potential is assumed, the distributed dark mass can 
have a core radius no larger than $\sim$5 pc. So, in either case, a density 
$\sim$10$^7$ \solmass\ pc$^{-3}$ is implied.

\begin{figure}
\hskip 0.85truein
\psfig{file=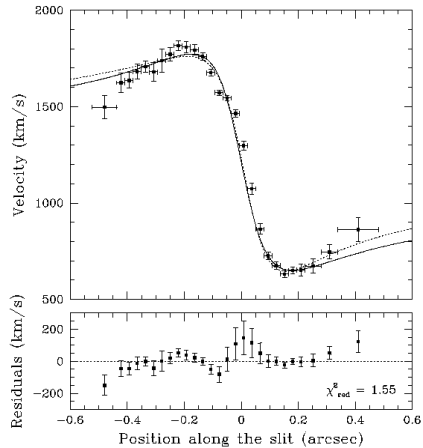,height=2.5truein,angle=0}
\caption{Optical emission-line rotation curve for the nuclear disk in M87. 
The two curves in the upper panel correspond to Keplerian thin disk models, 
and the bottom panel shows the residuals for one of the models 
(see Macchetto \etal 1997).} 
\end{figure}

The mildly active nucleus of NGC 4261 contains a rotating disk of dust and
ionized gas as well (Ferrarese, Ford, \& Jaffe 1996); like M87, the disk 
is slightly warped and shows traces of weak spiral structure.  
Although the FOS data for this object are rather noisy, they indicate 
that the gas largely undergoes circular motions.  The mass interior 
to $r$ = 15 pc is $M_{\rm {\small{MDO}}}$ = (4.9$\pm$1.0)\e{8} \solmass, and 
$M/L_V$ has an exceptionally high value of 2\e{3}.

The installation of the imaging spectrograph STIS in 1997 
at long last gives {\it HST} an efficient means to obtain spatially 
resolved spectra of the central regions of galaxies.  Much progress in the 
field is anticipated in the near future.  A taste of what might
be expected can be seen in the early-release observations of M84 
by Bower \etal (1998; Fig. 5).  M84 is almost a twin of M87 in terms of 
luminosity, and its central dark mass (1.5\e{9} \solmass), too, is similar.

\begin{figure}
\hskip 0.95truein
\psfig{file=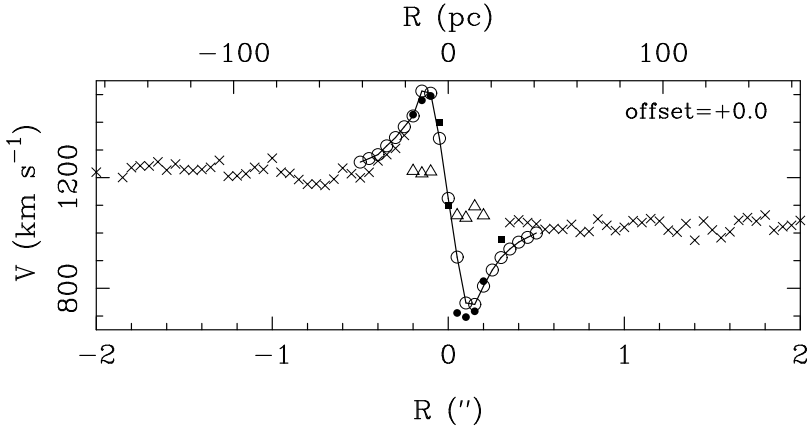,width=4.2truein,angle=0}
\caption{Optical emission-line rotation curve for the nuclear disk in M84 
obtained with STIS on {\it HST} (Bower \etal 1998).  The kinematics can be 
well fitted with a thin Keplerian disk model, which is plotted as 
open circles connected by the solid line.} 
\end{figure}

Lastly, I mention an interesting, unconventional case. The radio galaxy 
Arp 102B belongs to a minority of AGNs that display so-called double-peaked 
broad emission lines.  Several ideas have been proposed for the peculiar 
line profiles in this class of objects, but the favored explanation is that 
the lines originate from a relativistic accretion disk 
(Eracleous \etal 1997).  During the course of a long-term optical monitoring 
of Arp 102B, the intensity ratio of the two peaks of the H\al\ line 
displayed sinusoidal variations with a period of 2.2 years for several 
years (Newman \etal 1997).  The periodic signal was interpreted as 
arising from a ``hot spot'' in the accretion disk.  By modeling the 
line profile from the epochs when the hot spot was quiescent, one can 
estimate the radius and inclination angle of the spot's orbit, and, 
combined with its period, the enclosed mass.  The mass within $r$ = 0.005 pc 
turns out to be 2.2\e{8} \solmass, consistent with a moderately luminous 
($M_B\,\approx$ --20 mag) elliptical (see \S\ 7).

\subsection{Radio Spectroscopy of Water Masers}

Luminous 22-GHz emission from extragalactic water masers are preferentially 
detected in galaxies with active nuclei, where physical conditions, 
possibly realized in a circumnuclear disk (Claussen \& Lo 1986), evidently 
favor this form of maser emission.  With the detection in NGC 4258 of 
high-velocity features offset from the systemic velocity by $\sim \pm$900 
\kms\ (Nakai, Inoue, \& Miyoshi 1993), Watson \& Wallin (1994) already surmised 
that the maser spectrum of this Seyfert galaxy can be interpreted as arising 
from a thin Keplerian disk rapidly rotating around a mass of $\sim$10$^7$ 
\solmass. But the solid proof of this picture came from the high-resolution 
($\Delta\theta$ = 0\farcs0006$\times$0\farcs0003; $\Delta v$ = 0.2 \kms) VLBA 
observations of Miyoshi \etal (1995) who demonstrated that the maser spots 
trace a thin ($<$0.003 pc), nearly edge-on annulus with an inner radius 
of 0.13 pc and an outer radius of 0.26 pc.  The systemic features lie 
on the near side of the disk along the line-of-sight to the center (Fig. 6);
the high-velocity features delineate the edges of the disk on either side and 
follow a Keplerian rotation curve to very high accuracy (\lax 1\%).  The 
implied binding mass within 0.13 pc is 3.6\e{7} \solmass, which 
corresponds to a density of $>$4\e{9} \solmass\ pc$^{-3}$.   In fact, one can 
place a tighter constraint on the density.  The maximum deviation of the 
velocities from a Keplerian rotation curve limits the extent of the central 
mass to $r$ \lax 0.012 pc (Maoz 1995), from which follows that the density 
must be $>$5\e{12} \solmass\ pc$^{-3}$.   

\begin{figure}
\hskip 0.75truein
\psfig{file=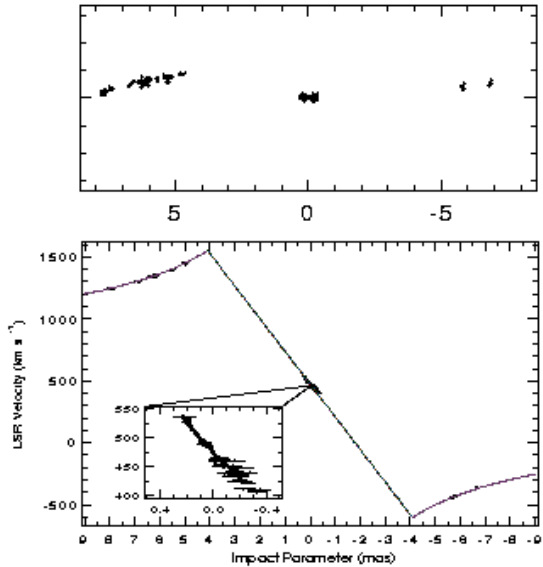,height=3.0truein,angle=0}
\caption{Water maser emission in NGC 4258 (Miyoshi et al. 1995).  {\it Top:} 
spatial distribution of the maser features; {\it bottom:} rotation curve.  
Adapted from Greenhill (1997).} 
\end{figure}

Two, possibly three, additional AGNs have H$_2$O megamasers suitable 
for tracing the central potential.  The spectrum of the maser source in the 
Seyfert nucleus of NGC 1068 also exhibits satellite features ($\pm$300 \kms) 
offset from the systemic velocity (Greenhill \etal 1996).   The redshifted and 
blueshifted emission again lie on a roughly linear, 2-parsec feature passing 
through the systemic emission (Greenhill 1998).  The rotation curve in this 
instance is sub-Keplerian, possibly because the disk has nonnegligible mass,
and the derived mass (1.7\e{7} \solmass\ within a radius of 0.65 pc) is less 
certain.

The maser in the nucleus of NGC 4945 shows a position and 
velocity distribution reminiscent of NGC 4258 as well: high-velocity features 
symmetrically straddle the systemic emission.  Greenhill, Moran, \& 
Herrnstein (1997) interpret the data, which in this case is 
considerably less accurate because of its location in the southern sky, 
in terms of an edge-on disk model and derive a central mass of 1.4\e{6} 
\solmass\ within $r$ = 0.3 pc.  This result is quite surprising because, 
as an Scd spiral, NGC 4945 is expected to be essentially bulgeless.  If the 
dark mass in its center is truly in the form of a SMBH, then SMBHs evidently 
can form without a bulge.

The H$_2$O megamaser source in NGC 3079 is potentially useful for mass 
determination.  Here, however, the complex spatial distribution 
of the emission regions and the large intrinsic widths of the lines 
complicate the analysis, and the interpretation of the data may not be unique.
Trotter \etal (1998) tentatively assign a central mass of 1\e{6} \solmass\ 
to this galaxy. 

\subsection{Determining Central Masses of Active Galactic Nuclei}

I mention one other method for determining masses in the central regions of
galaxies, specifically in AGNs.  Although AGNs largely provide the motivation
for searching for SMBHs, ironically it is precisely in these objects that 
conventional techniques used to measure masses fail.  The bright continuum 
emission of the active nucleus nearly always completely overpowers the stellar 
absorption lines near the center, and in many cases the narrow emission lines 
are significantly affected by nongravitational forces.

An approach taken in the past attempts to utilize the 
broad [(1--few)\e{3} \kms] emission lines that are thought to arise 
from the so-called broad-line region (BLR), a tiny, dense region
much less than a parsec from the central source.  Assuming that the line 
widths trace gravity, the mass follows from 
$\eta v^2 r_{\rm {\small{BLR}}}/G$, where $\eta\,\approx$ 1--3 depending on the 
kinematic model adopted.  The BLR radius has traditionally been estimated from 
photoionization arguments (e.g., Dibai 1981; Wandel \& Yahil 1985; 
Wandel \& Mushotzky 1986; Padovani, Burg, \& Edelson 1990), but recent 
variability studies indicate that the BLR is much more compact than previously 
thought (Netzer \& Peterson 1997).

\begin{figure}
\psfig{file=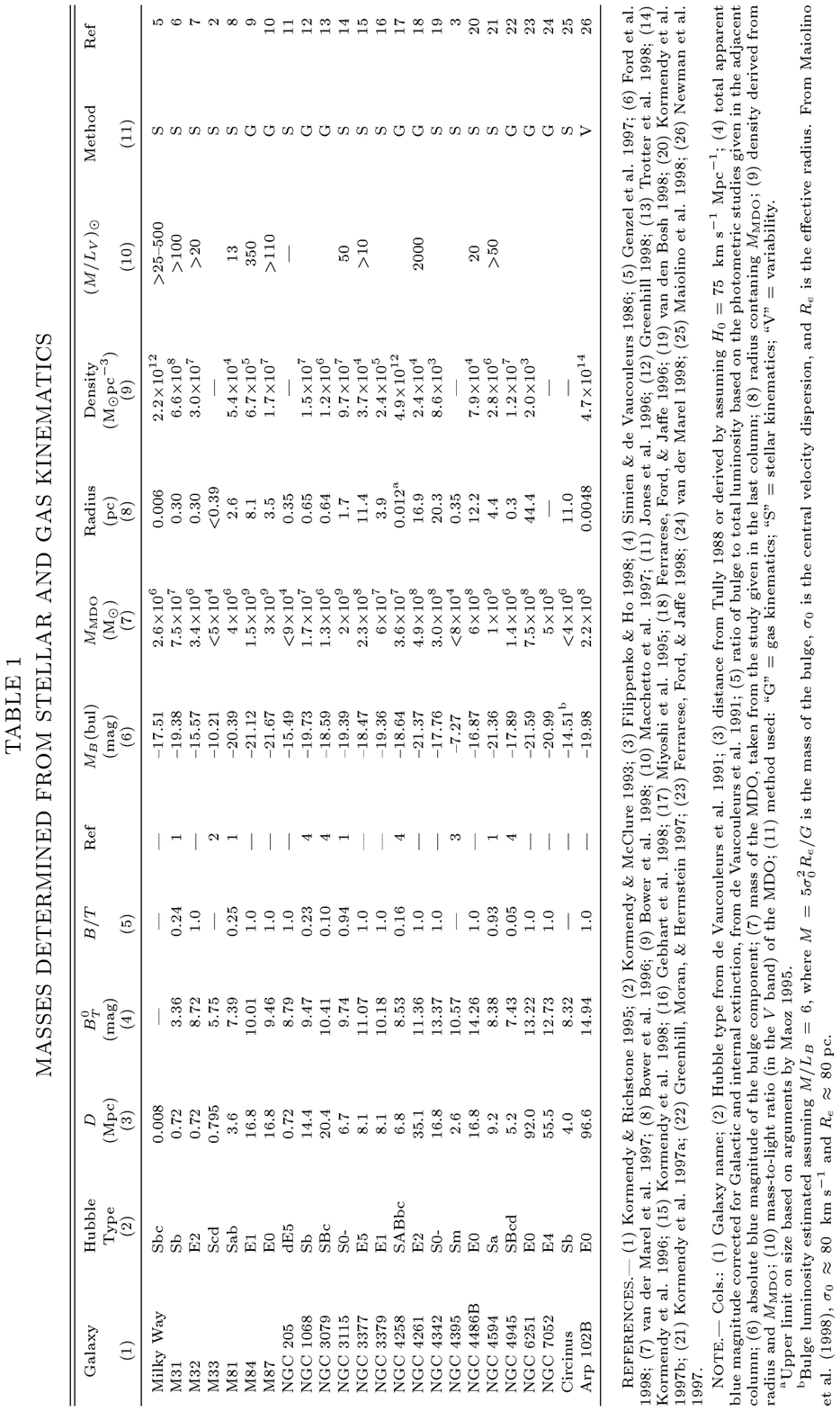,height=7.0truein,angle=180}
\end{figure}
\setcounter{figure}{6}
 
The continuum output from AGNs typically varies on
timescales ranging from days to months in the UV and optical bands.
Because the emission lines are predominantly photoionized by the central
continuum, they vary in response to the changes in the continuum, but with
a time delay (lag) that corresponds to the light-travel distance between the
continuum source and the line-emitting gas.
``Reverberation mapping'' (Blandford \& McKee 1982), therefore, in
principle allows one to estimate the luminosity-weighted radius of the BLR,
although in practice the complex geometry and ionization structure of the BLR
complicate the interpretation of the ``sizes'' derived by this method
(see Netzer \& Peterson 1997 for a recent review).
 
If the widths of the broad emission lines reflect bound gravitational motions,
as seems to be the case in most well-studied objects (Netzer \& Peterson
1997; but see Krolik 1997), then, adopting a reasonable kinematic model
(e.g., randomly moving clouds), the virial mass can be
estimated from $v^2 r_{\rm {\small{BLR}}}/G$.  If, instead, the clouds are 
infalling, as has been claimed in some cases, the mass will be smaller by a 
factor of 2.  One of the uncertainties in the application of this simple 
formalism lies in the choice of $v$.  What is appropriate?  One reasonable 
choice might be $v$ = ($\sqrt{3}/2$)FWHM, the full width at half-maximum of a 
representative broad line.  Yet another ambiguity is which line to use, since 
not all broad emission lines have the same widths.  Ultraviolet or 
high-ionization lines, for instance, generally have broader profiles than 
optical or low-ionization lines.  For the purposes of this exercise, I simply 
chose the line for which the most data exist (H\bet) in order to obtain 
as large a sample as possible.

\vskip 0.1truein 
\psfig{file=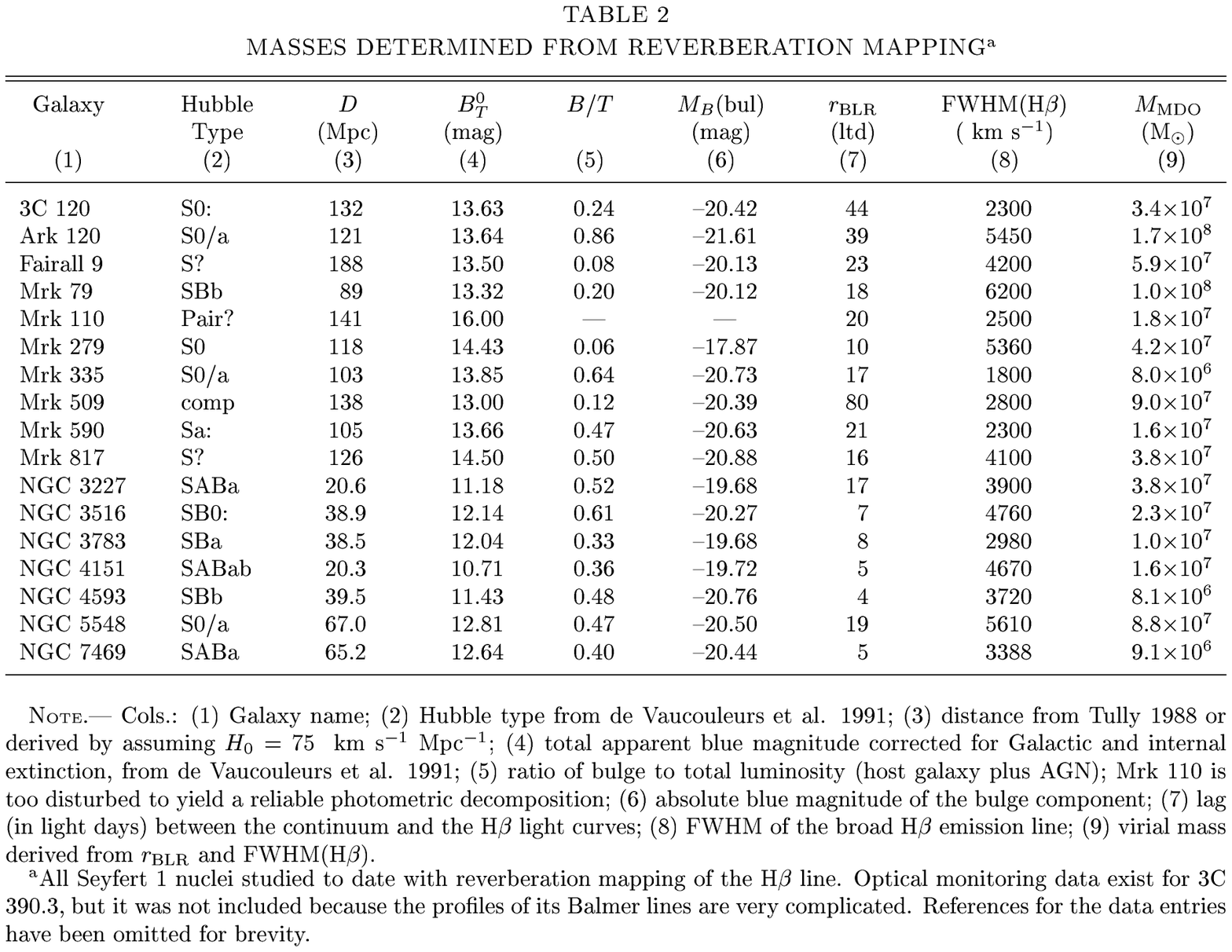,width=4.8truein,angle=0}

Table 2 lists the derived masses for the 17 Seyfert 1 galaxies that have been 
monitored extensively in the optical; eight of the objects appear in the 
compilation of Peterson \etal (1998).  Since the masses of MDOs
derived from gas and stellar kinematics show a loose correlation with the
bulge or spheroidal luminosity of the host galaxies (\S\ 7; Fig. 8{\it a}),
we can ask whether those derived from reverberation mapping follow such a
correlation.  I have estimated the $B$-band luminosities of the bulges of
the Seyferts based on published surface photometry of the host galaxies
(taking care to exclude the contribution of the AGN itself, which often can be
significant).  Figure 8{\it b} indicates that, at a fixed bulge luminosity,
the masses from reverberation mapping are {\it systematically lower} than the
masses obtained using conventional techniques, on average by about a factor
of 5.  It is encouraging that this admittedly crude method of mass estimation 
is not {\it too} far off the mark.  Notably, the scatter of $M_{\rm 
{\small{MDO}}}$ at a fixed luminosity is quite comparable in the two samples, 
and the constant offset suggests that one of the underlying assumptions 
in the mass estimate is incorrect.  Since the line width affects the mass
quadratically, it is conceivable that some measure of the line profile other
than the FWHM is more appropriate.

\section{Indirect, but Tantalizing Evidence}

Lastly, one additional piece of evidence, although it does not give a
direct measure of the central mass, cannot go unmentioned --- namely the
recent detection in AGNs of the broad iron K\al\ line at 6.4 keV.
This line has been known for some time to be a common feature in the hard
X-ray spectra of AGNs, and it is thought to arise from fluorescence of
the X-ray continuum off of cold material, presumably associated with the
accretion disk around the SMBH (e.g., Pounds \etal 1990).  The spectral
resolution of the existing data, however, was insufficient to
test the predicted line profile (Fabian \etal 1989).  The {\it ASCA}
satellite provided the much-awaited tell-tale signs in the deep exposure
of the Seyfert 1 galaxy MCG--6-30-15 (Tanaka \etal 1995): the Fe K\al\ line
exhibits Doppler motions that approach relativistic speeds ($\sim$100,000
\kms\ or 0.3$c$!) as well as an asymmetric red wing consistent with
gravitational redshift.  The best-fitting disk has an inner radius of
6 Schwarzschild radii.  The relativistic Fe K\al\ line, now seen
in a large number of sources (Nandra \etal 1997; Fig. 7), provides arguably 
the most compelling evidence to date for the existence of SMBHs.  Other 
mechanisms for generating the line profile are possible, but implausible 
(Fabian \etal 1995).   Detailed modeling of the line asymmetry has even the 
potential to measure the spin of the hole, but this is still very 
much a goal of the future given the current data quality and uncertainties in 
the modeling itself (e.g., Reynolds \& Begelman 1997; Rybicki \& Bromley 1998).

\begin{figure}
\hskip 0.5truein
\psfig{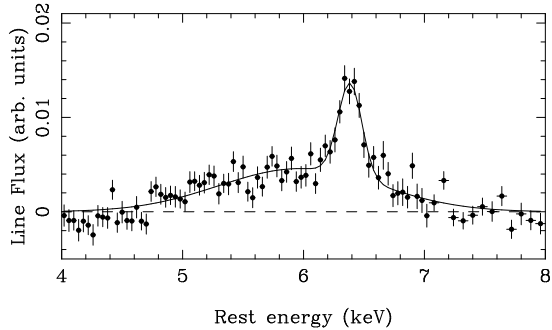}
\caption{The Fe K\al\ line in the composite spectrum of Seyfert 1 
nuclei (Nandra \etal 1997).  The solid line is a fit to the line profile 
using two Gaussians, a narrow component centered at 6.4 keV and a much
broader, redshifted component.} 
\end{figure}

\section{Are the Massive Dark Objects {\it Really} Black Holes?}

Thus far we have rigorously shown only that many galaxies contain central MDOs, 
not that the dark masses must be in the form of SMBHs.  Direct proof of 
the existence of SMBHs would require the detection of relativistic motions 
emanating from the vicinity of the Schwarzschild radius, 
$R_S\,=\,2GM_{\bullet}/c^2\,\approx\,10^{-5}(M_{\bullet}/10^8$ \solmass) pc.  
Even for our neighbor M31, $R_S$ subtends 3\e{-6} arcseconds, and the 
Galactic Center only a factor of 2 larger.  We are clearly still far from 
being able to achieve the requisite angular resolution and in the meantime 
must rely on indirect arguments.  

One approach seeks to identify some observational feature that might be 
taken as a fingerprint of the event horizon or of physical processes
uniquely associated with the environment of a BH.  One such
``signature'' might be the broad Fe K\al\ line discussed in \S\ 5; another 
is the high-energy power-law tail observed in some AGNs and Galactic BH
candidates (Titarchuk \& Zannias 1998).  And yet a third possibility is 
the advection of matter into the event horizon (Menou, Quataert, \& 
Narayan 1998).  

A different strategy appeals to the dynamical stability of the probable
alternative sources of the dark mass (Goodman \& Lee 1989; Richstone, Bower, 
\& Dressler 1990; van der Marel \etal 1997; Maoz 1998).  The absence of strong 
radial gradients in the stellar population, as measured by variations in color 
or spectral indices, implies that the large increase in $M/L$ toward the 
center cannot be attributed to a cluster of ordinary stars.  On the other 
hand, the underluminous mass could, in principle, be a cluster of stellar 
remnants (white dwarfs, neutron stars, and stellar-size BHs) or perhaps 
even substellar objects (planets and brown dwarfs).  To rule out these 
possibilities, however exotic they might seem,  one must show that the 
clusters cannot have survived over the age of the galaxy, and hence 
finding them would be highly improbable.

As most recently discussed by Maoz (1998), the two main processes that 
determine the lifetime of a star cluster are evaporation, whereby stars 
escape the cluster as a result of multiple weak gravitational scatterings, 
and physical collisions among the stars themselves. Exactly which dominates 
depends on the composition and size of the cluster, and its maximum possible 
lifetime can be computed for any given mass and density.  Maoz (1998) shows 
that in two galaxies, namely the Milky Way and NGC 4258, the density of the 
dark mass is so high (\gax$10^{12}$ \solmass\ pc$^{-3}$) that it cannot 
possibly be in the form of a stable cluster of stellar or substellar
remnants: their maximum ages [$\sim$(1--few)\e{8} yr] are much less than the 
ages of the galaxies.  The only remaining constituents allowed appear to be 
subsolar-mass BHs and elementary particles.  This constitutes very 
strong evidence that the MDOs --- at least in two cases --- are most likely 
SMBHs.   In the following discussion, I will adopt the simplifying 
viewpoint that all MDOs are SMBHs, bearing in mind that at the current 
resolution limit we cannot yet disprove the dark-cluster hypothesis for the 
majority of the objects.  

\section{The Black-Hole Mass/Bulge Mass Relation}

Does $M_{\bullet}$ depend at all on other properties of the host galaxies?  A 
much-discussed possibility is that $M_{\bullet}$ scales with the mass of 
the spheroidal component of the host (Kormendy 1993; KR; Faber \etal 1997; 
Magorrian \etal 1998; Richstone 1998; Ford \etal 1998; van der Marel 1998). 
The significance of the scatter in the correlation, or whether any correlation 
exists at all, is not yet certain.  It is somewhat disconcerting that 
different authors plotting the same objects do not always arrive at the 
same conclusion.  The discrepancies can often be traced to different 
assumptions about distances, source of bulge-to-disk decomposition, and 
even apparent magnitudes adopted for the host galaxies (e.g., extinction is 
not always corrected).  The set of host galaxy parameters I adopt is
compiled in Table 1.  

\begin{figure}
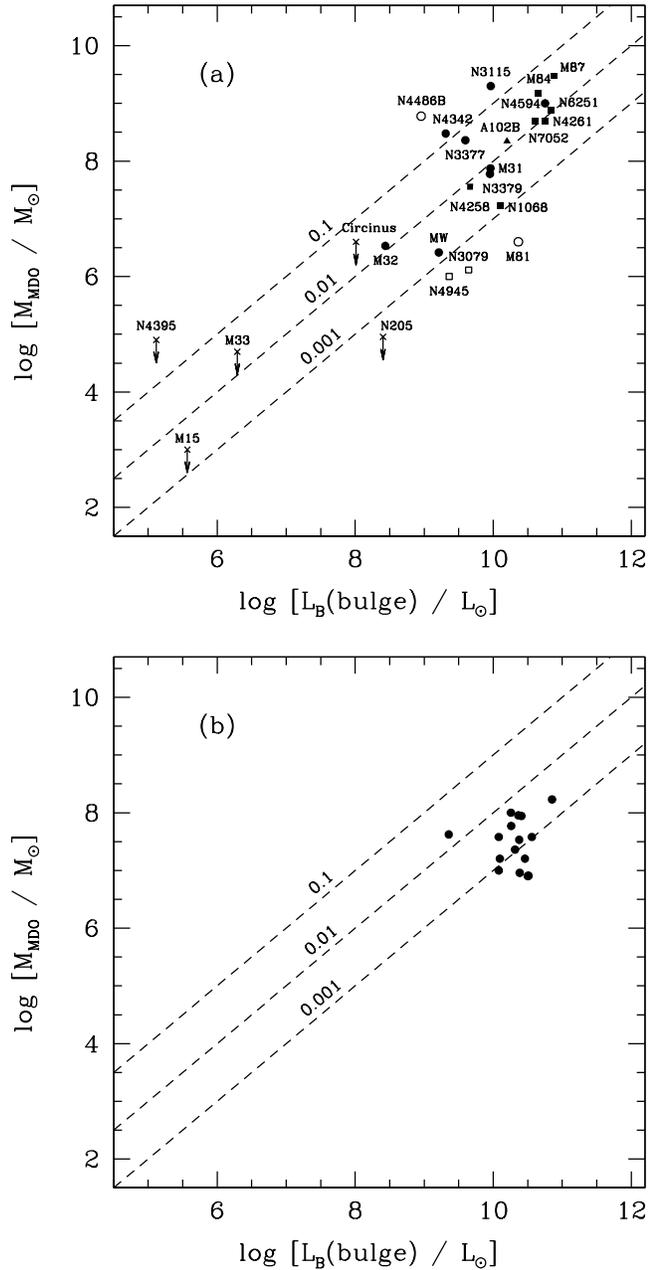

\vbox{
\vskip -0.25truein
\hskip 0.65truein
\psfig{file=Mbh_Lbul.ps1,height=3.8truein,angle=0}

\vskip -0.4truein
\hskip 0.65truein
\psfig{file=Mbh_Lbul.ps2,height=3.8truein,angle=0}
}
\caption{({\it a}) Log $M_{\bullet}$ versus log $L_B$(bulge) for the 
objects listed in Table 1.  The typical uncertainty of $M_{\bullet}$ is 
probably about a factor of 2.  Open symbols denote points that may have an 
exceptionally large uncertainty in either of the two variables (see text).  
Masses derived from stellar kinematics are plotted as {\it circles}, those 
from gas kinematics as {\it squares}, the unconventional case of Arp 102B as a 
{\it triangle}, and five upper limits as {\it crosses}.  Lines of constant 
mass to luminosity ratio are also shown.  ({\it b}) Same as in ({\it a}), but 
for the Seyfert galaxies listed in Table 2 (except Mrk 110).}
\end{figure}

Figure 8{\it a} illustrates that there indeed appears to be a trend of 
$M_{\bullet}$ increasing with bulge mass (luminosity).  It is encouraging to 
note that the central masses derived from gas and stellar kinematics do not 
show any obvious systematic offsets relative to one another.  
No obvious differentiation by Hubble type is evident either.  As has been 
noted by others, the scatter of $M_{\bullet}$ at a given luminosity is 
considerable, at least a factor of 10, perhaps up to 100.   The scatter 
may have been exacerbated slightly by four possibly anomalous points.  
NGC 4486B is a companion to M87, and it appears to have been tidally 
truncated; its original luminosity was probably higher.  On the other hand, 
the bulge luminosity of NGC 4945 could very well have been overestimated.  Its 
bulge-to-disk ratio was found using the relation of Simien \& de Vaucouleurs 
(1986), which may be inappropriate for a galaxy of such late Hubble type 
(Scd).  Finally, the masses of M81 and NGC 3079 are quite uncertain and 
probably have been underestimated.

The trend is much more significant when five upper limits are included.  
NGC 205, a dwarf elliptical companion of M31, contains a blue, compact nucleus 
with characteristics resembling an intermediate-age globular cluster. 
Its core radius, determined from {\it HST} photometry, combined with a 
ground-based measurement of its velocity dispersion yields an upper 
limit of 9\e{4} \solmass\ for any dark mass (Jones \etal 1996).  The bulgeless, 
late-type (Scd) spiral M33 also has a stringent upper limit on its central 
mass.  Its nuclear cluster is extremely tiny (core radius \lax 0.39 pc), and 
its central velocity dispersion is 21 \kms; Kormendy \& McClure (1993) put an 
upper limit of $M_{\bullet}\,\leq$ 5\e{4} \solmass.   NGC 4395 in 
some ways resembles M33, but it is even more extreme.  The nucleus is 
optically classified as a type 1.8 Seyfert (broad H\al\ and H\bet\ present), 
emits a largely nonstellar featureless continuum that extends into the 
UV (Filippenko, Ho, \& Sargent 1993), and displays variable 
soft X-ray emission and a compact flat-spectrum radio core (Moran \etal 
1998).  These properties alone would be unremarkable were it not for the 
fact that the nucleus has an absolute blue magnitude of only --9.8 and lives 
in a Magellanic spiral 2.6 Mpc away!  Filippenko \& Ho (1998) detected the 
Ca~II infrared triplet lines in absorption from echelle spectra taken with the 
Keck telescope, from which they were able to estimate the strength of the 
stellar component contributing to the nuclear light ($M_B$ = --7.3 mag) and 
the central stellar velocity dispersion ($\sigma\,\approx$ 30 \kms).  
Combining the velocity dispersion with a cluster size ($r$ \lax 0.7 pc) 
obtained from {\it HST} images, Filippenko \& Ho limit the central mass to 
\lax 8\e{4} \solmass.  The Circinus galaxy is thought to house a Seyfert 
nucleus, and if it contains a SMBH, its mass within $r\,\approx\,10$ pc has 
been constrained to be \lax 4\e{6} \solmass\ (Maiolino \etal 1998).  The last 
upper limit shown in the figure pertains to the globular cluster M15; 
following KR, I adopt an upper limit of $M_{\bullet}$ = 1\e{3} 
\solmass. 

However, before reading too much into this diagram, we should ask whether
the apparent correlation might arise from selection effects.
The absence of points on the upper left-hand corner is probably real;
there is nothing preventing us from detecting a massive BH in a
small galaxy.  Yet, we should be cautious, because very few low-mass
galaxies have been studied so far, most of the effort having been
focused on luminous, early-type systems.  On the other hand, the empty
region on the lower right-hand corner could be an artifact.  Small 
masses are difficult to detect at large distances, and most luminous 
galaxies are far away.  So the apparent correlation {\it could} be
an upper envelope.  Future observations are needed to settle this issue.


The median value of $M_{\bullet}/L_B({\rm bul})$ for the 20 detected 
objects is 0.012, which translates into a mass ratio of 0.002 for 
$M/L_B\,\approx\,6$ typical for old stellar populations (van der Marel 1991).
That is, on average about 0.2\% of the bulge mass is locked up in the form 
of a SMBH.  Magorrian \etal (1998) constructed axisymmetric $f(E,L_z)$ 
models for a sample of 32 early-type (mostly E and S0) galaxies having both 
{\it HST} photometry and ground-based stellar kinematics data, and they 
concluded that the data are consistent with nearly all of the galaxies 
having SMBHs.  The 29 detected objects have a median 
$M_{\bullet}/M_{\rm bul}\,\approx$ 0.005, higher than found here.  However, as 
Magorrian et al. realize, the assumption of a two-integral distribution 
function may have caused them to overestimate $M_{\bullet}$ (cf. van der 
Marel 1998).  Interestingly, quasars possibly also obey a similar 
$M_{\bullet}$-$M_{\rm bul}$ relation.  McLeod (1998) finds that, for the most 
luminous quasars, there exists a minimum host luminosity that increases 
with nuclear power.  Assuming that the quasar luminosities correspond to 
energy generation at the Eddington rate, $M_{\bullet}/M_{\rm bul}$ 
is again $\sim$0.002 (McLeod 1998).

With regard to the dead quasar prediction discussed in \S\ 1, recall that 
we expect to find on average a 10$^7$ \solmass\ BH for every 
$L_B\,\approx\,10^{10}$ \solum\ galaxy, or $M_{\bullet}/L_B({\rm bul})\, 
\approx$ 3.3\e{-3} \solmass/\solum\ since bulges contribute typical 30\% of 
the galaxy light in $B$ (Schechter \& Dressler 1987).  Evidently, if 
$\epsilon$ = 0.1, we have already found about three times that value.  
This implies that either $\epsilon$ is smaller than 0.1, or that quasars do 
not make up all of the AGN population.  

\section{Are Supermassive Black Holes Ubiquitous?}

They certainly have not been found in every case that has been looked.  
Kormendy has undertaken a systematic survey of a modest sample of galaxies 
(E--Sb), and his detection rate has been about 20\%\ (KR).  But, of 
course, many factors conspire against the detection of MDOs, and this estimate 
should be regarded as firm lower limit.  If one takes seriously the 
$M_{\bullet}$-$M_{\rm bul}$ relation described above, it is possible that 
{\it every} bulge contains a SMBH with an appropriately scaled size.  This 
view is supported by the statistical analysis of Magorrian \etal (1998).  In 
fact, the detection of an MDO in NGC 4945 (\S\ 4.2) and the presence of a 
{\it bona fide} AGN in NGC 4395 indicate that perhaps even some galaxies 
without bulges may have SMBHs.  

\begin{figure}   
\hskip 0.10truein
\psfig{file=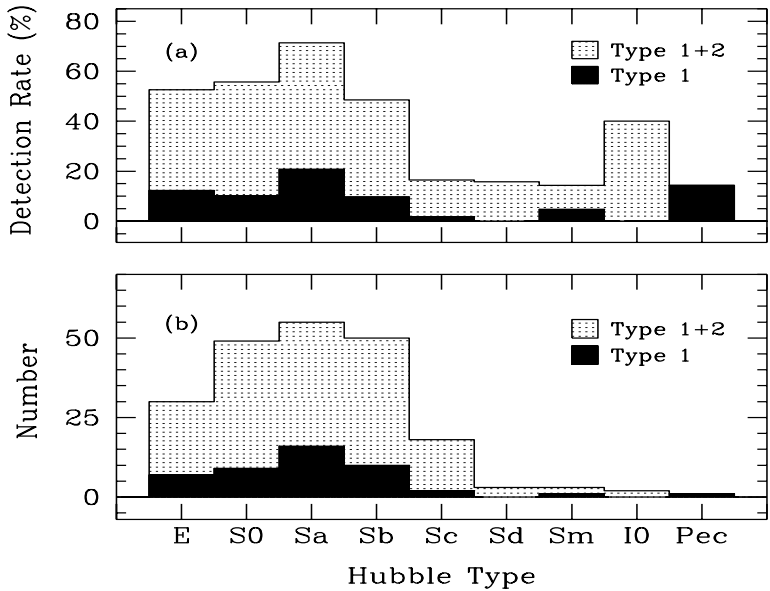,height=3.3truein,width=3.5truein,angle=0}
\caption{({\it a}) Detection rate and ({\it b}) number distribution of AGNs
as a function of Hubble type in the spectroscopic survey of Ho \etal (1995, 
1997).  ``Type 1'' AGNs (those with broad H$\alpha$) are shown separately 
from the total population (types 1 and 2).}
\end{figure}

Additional support for this picture comes from the growing evidence that 
nonstellar nuclear activity is very common in galaxies, much more so than 
conventionally believed based on the statistics of bright AGNs and quasars.
A recent spectroscopic survey of a large, statistically complete sample of 
nearby galaxies finds that over 40\% of all bright ($B_T\,\leq\,12.5$ mag) 
galaxies have nuclei that can be classified as ``active,'' and the percentage 
is even higher among early-type systems (E--Sbc), approaching 50\%--75\%\ 
(Ho, Filippenko, \& Sargent 1997).  Most of the nearby AGNs have much 
lower luminosities than traditionally studied active galaxies, and a 
greater heterogeneity in spectral types is found (low-ionization nuclei, or 
LINERs, are common, for example), but the evidence is overwhelming that many 
of these nuclei are truly accretion-powered sources (see 
Filippenko 1996; Ho et al. 1997).  Moreover, intrinsically weak, compact 
radio cores are known to be present in a significant fraction of elliptical 
and S0 galaxies (Sadler, Jenkins, \& Kotanyi 1989; Wrobel \& Heeschen 1991), 
almost all of which spectroscopically qualify as AGNs (Ho 1998).  

Within the conventional AGN paradigm, the observed widespread nuclear activity
implies that SMBHs are a generic component of many, perhaps most, 
present-day bulge-dominated galaxies, consistent with the picture emerging 
from the kinematic studies.  This is a remarkable statement.  It implies 
that SMBHs should not be regarded as ``freaks of nature'' that 
exist in only a handful of galaxies; rather, they must be accepted and 
understood as a normal component of galactic structure, one that arises 
naturally in the course of galaxy formation and evolution.

\section{Some Implications and Future Directions}

The SMBH hunting game is rapidly becoming a rather mature subject.  I think 
we have progressed from the era of ``the thrill of discovery'' to a point 
where we are on the verge of using SMBHs as astrophysical tools.  In this 
spirit, let me remark on a few of the ramifications of the existing 
observations and point out some of the more urgent directions that should
be pursued.

\noindent{\underbar{\it{A. The $M_{\bullet}$-$M_{\rm bul}$ relation.}}}
The apparent correlation between the mass of the central BH and the 
mass of the bulge, if borne out by future scrutiny, has significant implications
(see below).  From an observational point of view, the highest 
immediate priority is to populate the $M_{\bullet}$-$M_{\rm bul}$ 
diagram with objects spanning a wide range in luminosity, with the eventual 
aim of deriving a mass function for SMBHs.  The samples should 
be chosen with the following questions in mind.  (1) Is the apparent trend a 
true correlation or does it instead trace an upper envelope?  (2) If the 
relation is real, is it linear?  (3) What is the magnitude of the intrinsic 
scatter?  And (4) is there a minimum bulge luminosity (mass) below which SMBHs 
do not exist?

In the near future, the most efficient way to obtain mass measurements for 
relatively large numbers of galaxies is to exploit the capabilities of STIS 
on {\it HST}.  Several large programs are in progress.  Although VLBI 
spectroscopy of H$_2$O masers delivers much higher angular resolution, this 
technique is limited by the availability of suitably bright sources.  Conditions
which promote H$_2$O megamaser emission evidently are realized in only a tiny 
fraction of galaxies (Braatz, Wilson, \& Henkel 1996).

\noindent{\underbar{\it{B. The formation of SMBHs.}}}
The $M_{\bullet}$-$M_{\rm bul}$ relation offers some clues to the 
formation mechanism of SMBHs.  How does a galaxy know how to extract a 
constant, or at least a limiting, fraction of its bulge mass into a SMBH? 
An attractive possibility is by the normal dynamical evolution of the
galaxy core itself.   The spheroidal component of nearby galaxies can attain 
very high central stellar densities --- up to 10$^5$ \solmass\ pc$^{-3}$ 
(Faber \etal 1997) --- and some with distinct nuclei have even higher 
concentrations still (Lauer \etal 1995).  Although most galaxy cores are 
unlikely to have experienced dynamical collapse (Kormendy 1988c), the 
{\it innermost} regions have much shorter relaxation times, especially when 
considering a realistic stellar mass spectrum because the segregation of the 
most massive stars toward the center greatly accelerates the dynamical 
evolution of the system.  Lee (1995, and these proceedings) shows that, under 
conditions typical of galactic nuclei, core collapse and merging of 
stellar-size BHs can easily form a seed BH of moderate mass. 
Alternatively, the seed object may form via
the catastrophic collapse of a relativistic cluster of compact remnants 
(Quinlan \& Shapiro 1990).  In either case, subsequent accretion of gas and 
stars will augment the central mass, and, over a Hubble time, may produce the 
distribution of masses observed.  It is far too premature to tell whether 
SMBHs form through the secular evolution of galaxies, as suggested here, 
through processes associated with the initial formation of galaxies 
(e.g., Rees 1984, 1998; Silk \& Rees 1998), or both, depending on the galaxy 
type (elliptical vs. spiral galaxies).  But the stage is set for a serious 
discussion.  More sophisticated modeling of the growth of SMBHs that take into 
account a wider range of initial conditions in galactic nuclei (e.g., relaxing 
the adiabatic assumption or adopting more realistic density profiles for the 
stellar distribution) may eventually yield testable predictions (see, e.g., 
Stiavelli 1998).

\noindent{\underbar{\it{C. Influence of SMBHs on galactic structure.}}}
Norman, May, \& Van Albada 
(1985) showed through $N$-body simulations that a massive singularity in 
the center of a triaxial galaxy destroys the box-like stellar orbits and hence 
can erase the nonaxisymmetry, at least on small scales.  This has several 
important consequences.  First, it implies that the presence of a SMBH 
can influence the {\it global} structure and dynamics of galaxies.  
Second, the secular evolution of the axisymmetry of the central potential 
points to a natural mechanism for galaxies to self-regulate the transfer of 
angular momentum of the gas from large to small scales.  This negative-feedback 
process may limit the growth of the central BH {\it and} the 
accretion rate onto it, and hence may serve as a promising framework for 
understanding the physical evolution of AGNs.  Merritt \& Quinlan (1998) 
find that the timescale for effecting the transition from triaxiality 
to axisymmetry depends strongly on the fractional mass of the BH; 
the evolution occurs rapidly when $M_{\bullet}$/$M_{\rm bul}$ \gax 2.5\%, 
remarkably close to the observational upper limit (Fig. 8{\it a}).

\noindent{\underbar{\it{D. The origin of central cores.}}}  It is not 
understood why giant ellipticals have such shallow central light profiles.  
``Cores'' do not develop naturally in popular scenarios of structure 
formation, and even if they form, they are difficult to maintain against the 
subsequent acquisition of the dense, central regions of satellite galaxies 
that get accreted (Faber \etal 1997).  Moreover, the very presence of a SMBH,
whether it grew adiabatically in a preexisting stellar system or the 
galaxy formed by violent relaxation around it (Stiavelli 1998), ought to 
imprint a more sharply cusped light profile (see \S\ 2) than is observed.  
An intriguing possibility is that cores were created as a result of mergers, 
where one or more of the galaxies contains a BH.  As the single (Nakano \& 
Makino 1998) or binary (Makino 1997; Quinlan \& Hernquist 1997) BH sinks 
toward the center of the remnant due to dynamical friction, it heats the 
stars, thereby producing a ``fluffy'' core.  If this interpretation is 
correct, it would provide a simple, powerful tool to diagnose the formation 
history of galaxies.  

\noindent{\underbar{\it{E. Why are the black holes so black?}}}
It has been somewhat puzzling how the BHs can remain so
dormant.  No doubt the dwindled gas supply in the present epoch, especially
in ellipticals, is largely responsible for the inactivity.  Yet the accretion
rate cannot be zero; even in the absence of inflow from the general
interstellar medium, some gas is shed through normal mass loss from the
innermost evolved stars, and occasionally such stars get tidally disrupted 
(see below).  If
SMBHs are indeed present, the radiative efficiency of the accretion flow must
be orders of magnitude lower than that of ``standard'' optically thick,
geometrically thin disks.  Such a situation may be realized in accretion flows
where advection becomes important when the accretion rate is highly
sub-Eddington (Narayan \& Yi 1995; Abramowicz \etal 1995; Nakamura \etal 1996;
Chakrabarti 1996).  Sgr A$^*$ at the Galactic Center has a bolometric
luminosity of only $\sim$$10^{37}$ \lum, or $L_{\rm bol}/L_{\rm Edd}$
$\approx$ 3\e{-8} (Narayan, Yi, \& Mahadevan 1995); in the case of the
LINER nucleus of M81 (Ho, Filippenko, \& Sargent 1996), $L_{\rm bol}\,\approx\,
10^{41}$ \lum\ and $L_{\rm bol}/L_{\rm Edd}$ \lax $10^{-4}$.  The spectral
energy distributions emitted by both objects differ dramatically from those of
luminous AGNs and can be approximately matched by advective-disk models.
 
\noindent{\underbar{\it{F. Tidal disruption of stars.}}} 
The prevalence of SMBHs suggested by the existing evidence predicts a 
relatively high incidence of tidal disruptions of stars as they scatter into 
nearly radial orbits whose pericenters pass within the tidal radius of the 
BH (Rees 1998, and references therein).  For a typical stellar 
density of 10$^5$ stars pc$^{-3}$, $M_{\bullet}$ = 10$^6$--10$^8$ \solmass, 
and $\sigma$ = 100--300 \kms, a solar-type star will be disrupted once 
every 10$^2$--10$^4$ years.  (BHs more massive than 10$^8$ \solmass\ 
will swallow the star whole.)  Roughly half the debris becomes unbound and 
half gets captured into an accretion disk which undergoes a bright flare 
($\sim 10^{10}$ \solum) lasting a few months to a year.  The spectrum is 
expected to be mainly thermal and to peak in the extreme-UV and soft X-rays.  
The contribution to the near-UV and optical bands is uncertain; it depends on 
assumptions concerning the geometry of the accretion disk (thick or thin) and 
on whether an optically thick envelope can form.  For plausible parameters, 
Ulmer (1998) estimates that a 10$^7$ \solmass\ BH will produce a flare 
with an absolute magnitude of about --20 in $U$ and --18.5 in $V$.
The realization that SMBHs may be even more common than previously thought 
provides fresh motivation to search for such stellar flares; some observational 
strategies are mentioned by Rees (1998).  Here, I wish to stress that 
quantifying the rate of stellar disruptions can be used as a tool to study 
the demography of SMBHs out to relatively large distances and hence should be 
regarded as complementary to the kinematic searches.

\acknowledgements {
I am grateful to S.~K. Chakrabarti for the invitation to participate in this 
workshop and his help in arranging a pleasurable visit to India.  I thank 
G.~A. Bower, S. Collier, R. Genzel, L.~J. Greenhill, J. Kormendy, R. Maiolino, 
D. Maoz, E. Maoz, and K. Nandra  for contributing to, or for providing 
comments that have improved the presentation of, the material in this paper.  
This work was supported by a postdoctoral fellowship from the 
Harvard-Smithsonian Center for Astrophysics and by NASA grants 
from the Space Telescope Science Institute (operated by AURA, Inc., under 
NASA contract NAS 5-26555). }

\bigskip
\centerline{\bf{References}}

\refindent
Abramowicz, M.~A., Chen, X., Kato, S., Lasota, J.-P., \& Regev, O. 1995,
\apj, 438, L37

\refindent
Bahcall, J.~N., \& Wolf, R.~A. 1976, \apj, 209, 214

\refindent
Bender, R., Burstein, D., \& Faber, S.~M. 1992, \apj, 399, 462


\refindent
Binney, J., \& Mamon, G.~A. 1982, \mnras, 200, 361

\refindent
Blandford, R.~D., \& McKee, C.~F. 1982, \apj, 255, 419

\refindent
Blandford, R.~D., \& Rees, M.~J. 1992, in Testing the AGN Paradigm, ed. 
S. Holt, S. Neff, \& M.  Urry (New York: AIP), 3

\refindent
Bower, G.~A., \etal 1998, \apj, 492, L111

\refindent
Bower, G.~A., Wilson, A.~S., Heckman, T.~M., \& Richstone, D.~O. 1996, in 
The Physics of LINERs in View of Recent Observations, ed. M. Eracleous et al.
(San Francisco: ASP), 163

\refindent
Braatz, J.~A., Wilson, A.~S., \& Henkel, C. 1996, \apjs, 106, 51



\refindent
Carollo, C.~M., Franx, M., Illingworth, G.~D., \& Forbes, D.~A. 1997, \apj,
481, 710


\refindent
Chakrabarti, S.~K. 1995, \apj, 441, 576

\refindent
Chakrabarti, S.~K. 1996, \apj, 464, 664

\refindent
Chokshi, A., \& Turner, E.~L. 1992, \mnras, 259, 421


\refindent
Claussen, M.~J., \& Lo, K.-Y. 1986, \apj, 308, 592


\refindent
Crane, P., \etal 1993, \aj, 106, 1371


\refindent
Cretton, N., de Zeeuw, P.~T., van der Marel, R.~P., \& Rix, H.-W. 1998, \apj,
in press

\refindent
Dehnen, W. 1995, \mnras, 274, 919


\refindent
de Vaucouleurs, G., de Vaucouleurs, A., Corwin, H.~G., Jr., Buta, R.~J.,
Paturel, G., \& Fouqu\'e, R. 1991, Third Reference Catalogue of Bright
Galaxies (New York: Springer) (RC3)

\refindent
Devereux, N.~A., Ford, H.~C., \& Jacoby, G. 1997, \apj, 481, L71


\refindent
Dibai, E.~A. 1981, Soviet Astron., 24, 389





\refindent
Dressler, A., \& Richstone, D.~O. 1988, \apj, 324, 701

\refindent
Dressler, A., \& Richstone, D.~O. 1990, \apj, 348, 120

\refindent
Duncan, M.~J., \& Wheeler, J.~C. 1980, \apj, 237, L27

\refindent
Eckart, A., \& Genzel, R. 1996, \nat, 383, 415
 
\refindent
Eckart, A., \& Genzel, R. 1997, \mnras, 284, 576


\refindent
Eracleous, M., Halpern, J.~P., Gilbert, A.~M., Newman, J.~A., \& Filippenko,
A.~V. 1997, \apj, 490, 216


\refindent
Faber, S.~M., \etal 1987, in Nearly Normal Galaxies, ed. S.~M. Faber (New York:
Springer), 175

\refindent
Faber, S.~M., \etal 1997, \aj, 114, 1771

\refindent
Fabian, A.~C., Nandra, K., Reynolds, C.~S., Brandt, W.~N., Otani, C., Tanaka,
Y., Inoue, H., \& Iwasawa, K. 1995, \mnras, 277, L11

\refindent
Fabian, A.~C., Rees, M.~J., Stella, L., \& White, N.~E. 1989, \mnras,
238, 729

\refindent
Ferrarese, L., Ford, H.~C., \& Jaffe, W. 1996, \apj, 470, 444

\refindent
Ferrarese, L., Ford, H.~C., \& Jaffe, W. 1998, \apj, in press


\refindent
Filippenko, A.~V. 1996, in The Physics of LINERs in View
of Recent Observations, ed.  M. Eracleous et al. (San Francisco: ASP), 17

\refindent
Filippenko, A.~V., \& Ho, L.~C. 1998, \apj, submitted

\refindent
Filippenko, A.~V., Ho, L.~C., \& Sargent, W.~L.~W. 1993, \apj, 410, L75

\refindent
Fillmore, J.~A., Boroson, T.~A., \& Dressler, A. 1986, \apj, 302, 208

\refindent
Ford, H.~C., \etal 1994, \apj, 435, L27

\refindent
Ford, H.~C., Tsvetanov, Z.~I., Ferrarese, L., \& Jaffe, W. 1998, in IAU Symp.
184, The Central Regions of the Galaxy and Galaxies, ed. Y. Sofue (Dordrecht: 
Kluwer), in press


\refindent
Gebhardt, K., \etal 1998, \aj, in press

\refindent
Genzel, R., Eckart, A., Ott, T., \& Eisenhauer, F. 1997, \mnras, 291, 219

\refindent
Genzel, R., Thatte, N., Krabbe, A., Kroker, H., \& Tacconi-Garman, L.~E. 1996,
\apj, 472, 153

\refindent
Gerhard, O.~E. 1993, \mnras, 265, 213

\refindent
Ghez, A.~M., \etal 1998, in IAU Symp. 184, The Central Regions of the
Galaxy and Galaxies, ed. Y. Sofue (Dordrecht: Kluwer), in press

\refindent
Goodman, J., \& Lee, H.~M. 1989, \apj, 337, 84


\refindent
Greenhill, L.~J. 1997, in IAU Colloq. 159, Emission Lines in Active Galaxies: 
New Methods and Techniques, ed. B.~M. Peterson, F.-Z. Cheng, \& A.~S. Wilson 
(San Francisco: ASP), 394

\refindent
Greenhill, L.~J. 1998, in IAU Colloq. 164, Radio Emission from Galactic and 
Extragalactic Compact Sources, ed. A. Zensus, G. Taylor, \& J. Wrobel (San 
Francisco: ASP), in press

\refindent
Greenhill, L.~J., Gwinn, C.~R., Antonucci, R., \& Barvainis, R. 1996, \apj,
472, L21

\refindent
Greenhill, L.~J., Moran, J.~M., \& Herrnstein, J.~R. 1997, \apj, 481, L23



\refindent
Harms, R.~J., \etal 1994, \apj, 435, L35



\refindent
Ho, L.~C. 1998, \apj, submitted

\refindent
Ho, L.~C., \etal 1998, in preparation

\refindent
Ho, L.~C., Filippenko, A.~V., \& Sargent, W.~L.~W. 1995, \apjs, 98, 477

\refindent
Ho, L.~C., Filippenko, A.~V., \& Sargent, W.~L.~W. 1996, \apj, 462, 183

\refindent
Ho, L.~C., Filippenko, A.~V., \& Sargent, W.~L.~W. 1997, \apj, 487, 568


\refindent
Jaffe, W., Ford, H.~C., Ferrarese, L., van den Bosch, F., \& O'Connell, R.~W.
1993, \nat, 364, 213

\refindent
Jaffe, W., \etal  1994, \aj, 108, 1567

\refindent
Jones, D. H., \etal 1996, \apj, 466, 742



\refindent
Kormendy, J. 1988a, \apj, 325, 128

\refindent
Kormendy, J. 1988b, \apj, 335, 40

\refindent
Kormendy, J. 1988c, in Supermassive Black Holes, ed. M. Kafatos (Cambridge:
Cambridge Univ. Press), 219

\refindent
Kormendy, J. 1993, in The Nearest Active Galaxies, ed. J. Beckman, L. Colina, 
\& H. Netzer (Madrid: CSIC Press), 197

\refindent
Kormendy, J., \etal 1996, \apj, 459, L57

\refindent
Kormendy, J., \etal 1997a, \apj, 473, L91

\refindent
Kormendy, J., \etal 1997b, \apj, 482, L139

\refindent
Kormendy, J., Bender, R., Evans, A.~S., \& Richstone, D. 1998, \aj, in press

\refindent
Kormendy, J., \& McClure, R.~D. 1993, \aj, 105, 1793


\refindent
Kormendy, J., \& Richstone, D.~O. 1995, \annrev, 33, 581 (KR)


\refindent
Krabbe, A., \etal 1995, \apj, 447, L95

\refindent
Krolik, J.~H. 1997, in  IAU Colloq. 159, Emission Lines in Active Galaxies: 
New Methods and Techniques, ed. B.~M. Peterson, F.-Z. Cheng, \& A.~S. Wilson 
(San Francisco: ASP), 459

\refindent
Lacy, J.~H., Townes, C.~H., Geballe, T.~R., \& Hollenbach, D.~J. 1980,
\apj, 241, 132

\refindent
Lauer, T.~R., \etal 1992, \aj, 103, 703

\refindent
Lauer, T.~R., \etal 1995, \aj, 110, 2622

\refindent
Lee, H.~M. 1995, \mnras, 272, 605


\refindent
Lin, H., \etal 1996, \apj, 464, 60




\refindent
Lynden-Bell, D. 1969, \nat, 223, 690

\refindent
Lynden-Bell, D., \& Rees, M.~J. 1971, \mnras, 152, 461

\refindent
Macchetto, F., Marconi, A., Axon, D.~J., Capetti, A., Sparks, W.~B., \&
Crane, P. 1997, \apj, 489, 579


\refindent
Magorrian, J., \etal 1998, \aj, in press

\refindent
Maiolino, R., Krabbe, A., Thatte, N., \& Genzel, R. 1998, \apj, 493, 650

\refindent
Makino, F. 1997, \apj, 478, 58



\refindent
Maoz, E. 1995, \apj, 447, L91

\refindent
Maoz, E. 1998, \apj, 494, L181


\refindent
McLeod, K.~K. 1998, in Quasar Hosts, ed. D. Clements \& I. Perez-Fournon
(Berlin: Springer-Verlag), in press

\refindent
Menou, K., Quataert, E., \& Narayan, R. 1998, in Proc. of the 8th Marcel
Grossmann Meeting on General Relativity (Jerusalem), in press


\refindent
Merritt, D., \& Quinlan, G. 1998, \apj, in press

\refindent
Miyoshi, M., Moran, J., Herrnstein, J., Greenhill, L., Nakai, N., Diamond, P., 
\& Inoue, M. 1995, \nat, 373, 127

\refindent
Moran, E.~C., Filippenko, A.~V., Ho, L.~C., Belloni, T., Shields,
J.~C., Snowden, S.~L., \& Sramek, R.~A. 1998, \apj, submitted

\refindent
Nakai, N., Inoue, M., \& Miyoshi, M. 1993, \nat, 361, 45

\refindent
Nakamura, K.~E., Matsumoto, R., Kusunose, M., \& Kato, S. 1996, PASJ, 48,
761

\refindent
Nakano, T., \& Makino, J. 1998, \apj, in press

\refindent
Nandra, K., George, I.~M., Mushotzky, R.~F., Turner, T.~J., \& Yaqoob, T.
1997, \apj, 477, 602

\refindent
Narayan, R., \& Yi, I. 1995, 452, 710

\refindent
Narayan, R., Yi, I., \& Mahadevan, R. 1995, \nat, 374, 623

\refindent
Netzer, H., \& Peterson, B.~M. 1997, in Astronomical Time Series, ed. 
D. Maoz, A. Sternberg, \& E.~M. Leibowitz (Dordrecht: Kluwer), 85

\refindent
Newman, J.~A., Eracleous, M., Filippenko, A.~V., \& Halpern, J.~P. 1997,
\apj, 485, 570

\refindent
Nieto, J.-L., Bender, R., Arnaud, J., \& Surma, P. 1991, \aa, 244, L2

\refindent
Norman, C.~A., May, A., \& Van Albada, T.~S. 1985, \apj, 296, 20

\refindent
Padovani, P., Burg, R., \& Edelson, R.~A. 1990, \apj, 353, 438

\refindent
Peebles, P.~J.~E. 1972, \apj, 178, 371





\refindent
Peterson, B.~M., Wanders, I., Bertram, R., Hunley, J.~F., Pogge, R.~W., \&
Wagner, R.~M. 1998, \apj, in press

\refindent
Pounds, K., Nandra, K., Stewart, G.~C., George, I.~M., \& Fabian, A.~C.
1990, \nat, 344, 132


\refindent
Qian, E.~E., de Zeeuw, P.~T., van der Marel, R.~P., \& Hunter, C. 1995, 
\mnras, 274, 602

\refindent
Quinlan, G.~D., \& Hernquist, L. 1997, New Astron., 2(6), 533

\refindent
Quinlan, G.~D., Hernquist, L., \& Sigurdsson, S. 1995, \apj, 440, 554


\refindent
Quinlan, G.~D., \& Shapiro, S.~L. 1990, \apj, 356, 483

\refindent
Rees, M.~J. 1984, \annrev, 22, 471



\refindent
Rees, M.~J. 1998, in Proc. Chandrasekhar Memorial Conf., Black Holes and
Relativity, ed. R. Wald (Chicago: Chicago Univ. Press), in press


\refindent
Reynolds, C.~S., \& Begelman, M.~C. 1997, \apj, 488, 109

\refindent
Richstone, D.~O. 1998, in IAU Symp. 184, The Central Regions of the Galaxy and 
Galaxies, ed. Y. Sofue (Dordrecht: Kluwer), in press

\refindent
Richstone, D.~O., Bower, G., \& Dressler, A. 1990, \apj, 353, 118

\refindent
Richstone, D.~O., \& Tremaine, S. 1984, \apj, 286, 27

\refindent
Richstone, D.~O., \& Tremaine, S. 1985, \apj, 296, 370

\refindent
Richstone, D.~O., \& Tremaine, S. 1988, \apj, 327, 82


\refindent
Rybicki, G.~B., \& Bromley, B.~C. 1998, \apj, in press

\refindent
Sadler, E.~M., Jenkins, C.~R., \& Kotanyi, C.~G. 1989, \mnras, 240, 591


\refindent
Salpeter, E.~E. 1964, \apj, 140, 796


\refindent
Sargent, W.~L.~W., Young, P.~J., Boksenberg, A., Shortridge, K., Lynds, C.~R., 
\& Hartwick, F.~D.~A. 1978, \apj, 221, 731

\refindent
Schechter, P.~L., \& Dressler, A. 1987, \aj, 94, 563


%
%
%
%
%
\refindent
Silk, J., \& Rees, M.~J 1998, \aa, 331, L1

\refindent
Simien, F., \& de Vaucouleurs, G. 1986, \apj, 302, 564

%
\refindent
Soltan, A. 1982, \mnras, 200, 115

\refindent
Stiavelli, M. 1998, \apj, 495, L91

 
 


\refindent
Tanaka, Y., \etal 1995, \nat, 375, 659

\refindent
Titarchuk, L., \& Zannias, T. 1998, \apj, 493, 863

 
\refindent
Trotter, A.~S., Greenhill, L.~J., Moran, J.~M., Reid, M.~J., Irwin, J.~A.,
\& Lo, K.-Y. 1998, \apj, 495, 740

\refindent
Tully, R.~B. 1988, Nearby Galaxies Catalog (Cambridge: Cambridge Univ. Press)

 
\refindent
Ulmer, A. 1998, \apj, in press

\refindent
van den Bosch, F.~C. 1998, in IAU Symp. 184, The Central Regions of the Galaxy 
and Galaxies, ed. Y. Sofue (Dordrecht: Kluwer), in press

\refindent
van den Bosch, F.~C., Jaffe, W., \& van der Marel, R.~P. 1998, \mnras, 293, 343

\refindent
van der Marel, R.~P. 1991, \mnras, 253, 710

\refindent
van der Marel, R.~P. 1994a, \mnras, 270, 271

\refindent
van der Marel, R.~P. 1994b, \apj, 432, L91

\refindent
van der Marel, R.~P.  1998, in IAU Symp. 186, Galaxy Interactions at Low
and High Redshift, ed. D.~B. Sanders \& J. Barnes (Dordrecht: Kluwer), in press

\refindent
van der Marel, R.~P., Cretton, N., de Zeeuw, P.~T., \& Rix, H.-W. 1998, \apj,
493, 613

\refindent
van der Marel, R.~P., de Zeeuw, P.~T., Rix, H.-W., \& Quinlan, G.~D.
1997, \nat, 385, 610

\refindent
van der Marel, R.~P., Evans, N.~W., Rix, H.-W., White, S.~D.~M., \& de Zeeuw,
P.~T. 1994a, \mnras, 271, 99

\refindent
van der Marel, R.~P., Rix, H.-W., Carter, D., Franx, M., White, S.~D.~M., \&
de Zeeuw, P.~T. 1994b, \mnras, 268, 521


\refindent
Wandel, A., \& Mushotzky, R.~F. 1986, \apj, 306, L61

\refindent
Wandel, A., \& Yahil, A. 1985, \apj, 295, L1


\refindent
Watson, W.~D., \& Wallin, B.~K. 1994, \apj, 432, L35

\refindent
Wrobel, J.~M., \& Heeschen, D.~S. 1991, \aj, 101, 148


\refindent
Young, P.~J. 1980, \apj, 242, 1232


\refindent
Young, P.~J., Westphal, J.~A., Kristian, J., Wilson, C.~P., \& Landauer, F.~P.
1978, \apj, 221, 721

\refindent
Zel'dovich, Ya.~B., \& Novikov, I.~D. 1964, Sov. Phys. Dokl., 158, 811

\end{document}